\documentclass[onecolumn,superscriptaddress,nobalancelastpage,12pt,pra,a4paper]{revtex4}
\usepackage{graphicx}
\usepackage{dcolumn}
\usepackage{bm}
\usepackage{verbatim}
\usepackage{soul}
\usepackage[dvipsnames]{xcolor}
\colorlet{RED}{red}
\usepackage{braket}
\usepackage{dsfont}
\usepackage{amsmath,subfigure,float}
\usepackage{url}
\usepackage{ctable}
\usepackage{soul}
\usepackage[USenglish]{babel}
\usepackage{titlesec}
\addto{\captionsUSenglish}{}

\newcommand{\avgp}[1]{\left\langle #1 \right\rangle}

\newcommand{\kk}{\mathbf{k}}

\titleformat{\section}{\normalfont\large\bfseries}{}{0pt}{}
\titleformat{\subsection}{\normalfont}{}{0pt}{\ul}
\titlespacing{\subsection}{0pt}{\parskip}{4pt}
\titleformat{\subsubsection}{\normalfont\large}{}{0pt}{}
\titlespacing{\subsubsection}{0pt}{\parskip}{4pt}

\begin{document}

\title{Experimental quantum imaging distillation \\ with undetected light}

\author{Jorge Fuenzalida,$^{1,2,\dagger,\star}$  Marta Gilaberte Basset,$^{1,3,\star}$ Sebastian T{\"o}pfer,$^{1,2,\star}$ Juan P. Torres,$^{4,\ddagger}$ and Markus Gr{\"a}fe$^{1,2,3,\S}$
\vspace{1em}
\\
\it $^{1}$Fraunhofer Institute for Applied Optics and Precision Engineering IOF, Albert-Einstein-Str. 7, 07745 Jena, Germany. \\ \vspace{0.5em}
\it $^{2}$Institute of Applied Physics, Technical University of Darmstadt, Schloßgartenstraße 7, 64289 Darmstadt, Germany.\\ 
\vspace{0.5em}
\it $^{3}$Friedrich Schiller University Jena, Abbe Center of Photonics,\\ 
\it Albert-Einstein-Str. 6, 07745 Jena, Germany.\\ \vspace{0.5em}
\it $^{4}$ICFO-Institut de Ciencies Fotoniques, The Barcelona Institute of Science and Technology,\\
\it 08860 Castelldefels, Spain, and Dept. Signal Theory and Communications, Universitat Politecnica de Catalunya, 08034 Barcelona, Spain.\\
\vspace{0.5em}
$^{\dagger}$ jorge.fuenzalida@tu-darmstadt.de\\
$^{\ddagger}$ juanp.torres@icfo.eu\\
$^{\S}$ markus.graefe@tu-darmstadt.de\\
$^{\star}$These authors contributed equally.
}

\date{\today}

\maketitle

\section{Abstract}
Imaging based on the induced coherence effect makes use of photon pairs to obtain information of an object without detecting the light that probes it. While one photon illuminates the object, only its partner is detected, so no measurement of coincidence events are needed. The sought-after object's information is revealed observing a certain interference pattern on the detected photon. Here we demonstrate experimentally that this imaging technique can be made resilient to noise. We introduce an imaging distillation approach based on the interferometric modulation of the signal of interest. We show that our scheme can generate a high-quality image of an object even against noise levels up to 250 times the actual signal of interest. We also include a detailed theoretical explanation of our findings.

\section{Introduction}
Quantum imaging~\cite{gilaberte2019perspectives} is an emerging and promising field in quantum technologies with certified advantages over classical protocols. This has been demonstrated in different scenarios: in schemes that work in the low-photon flux regime~\cite{aspden2013epr,morris2015imaging}, in schemes that make use of undetected probing photons~\cite{lemos:2014:quantum, gilaberte2021video}, for super-resolution imaging~\cite{schwartz2013superresolution,classen2017superresolution,unternahrer2018super,tenne2019super}, sub-shot-noise imaging~\cite{brida2010experimental,sabines2019twin,casacio2021quantum}, or enhanced two-photon absorption processes~\cite{teich1997entangled}. Moreover, protocols in quantum imaging with no classical counterpart have been developed based on quantum interference~\cite{ndagano2022quantum}, and entanglement~\cite{defienne2021polarization,camphausen2021quantum}. In recent years, it has also been proven that quantum imaging protocols can be resilient to noise~\cite{lloyd2008enhanced,tan2008quantum,lopaeva2013experimental}.

\vspace{5mm}
Distillation (also known as purification) is the process wherein the decoherence introduced in a quantum system by the environment can be removed~\cite{pan2001entanglement}. In quantum imaging, the effect of the environment can be modeled through classical illumination superimposed over a quantum image on the camera. Since most cameras only detect intensity, quantum and classical images seem to be indistinguishable. However, quantum correlations of photon pairs can be employed to differentiate the quantum image from a classical one. Quantum imaging distillation has been implemented with one and several photon pair degrees of freedom~\cite{duncan2019quantum,defienne2019quantum,zhang2020multidimensional,gregory2020imaging,gregory2021noise,zhao2022light}. To the best of our knowledge, every implementation to date has used the joint detection of photon pairs. In this work, we introduce and experimentally verify a quantum imaging distillation technique that employs the detection of single photons only. 

\vspace{5mm}
Quantum imaging with undetected light (QIUL)~\cite{lemos:2014:quantum,lahiri2015theory,fuenzalida2022resolution,vega2022fundamental} is a two-photon wide-field interferometric imaging technique. In QIUL, one photon illuminates an object and its partner photon is detected on the camera. The photon that illuminates the object remains undetected. Using an interferometric configuration, the object information is transferred to the detected photon interference pattern. Due to its unique detection advantage, QIUL has been employed to probe samples with unconventional wavelengths while visible light is detected~\cite{kviatkovsky2020microscopy,paterova2020hyperspectral,paterova2020quantum,paterova2021non,haase2023phase}. Up to date, the effects of noise in QIUL have not yet been explored. 

\vspace{5mm}
Here we introduce a source of noise in a QIUL scheme and study the resilience of the quantum imaging technique. The properties of the noise, its intensity and variance, are changed during this study. We perform a quantum imaging distillation technique based on quantum phase-shifting digital holography~\cite{topfer2022quantum}. Our distillation technique uses phase modulation of the undetected photon to vary the interference pattern detected on the camera. We notice that if the intensity difference of the interference patterns is bigger than the variance of the noise, the quantum image can be distilled. We also observe that the noise variance affects the distilled quantum images linearly in their phase estimation. Our technique shows a good performance, even for noise intensities above 250 times the quantum signal intensity. 

\section*{Results}

\subsection*{Distillation principle}
\begin{figure}[bt]
	\centering
\includegraphics[width=0.8\linewidth]{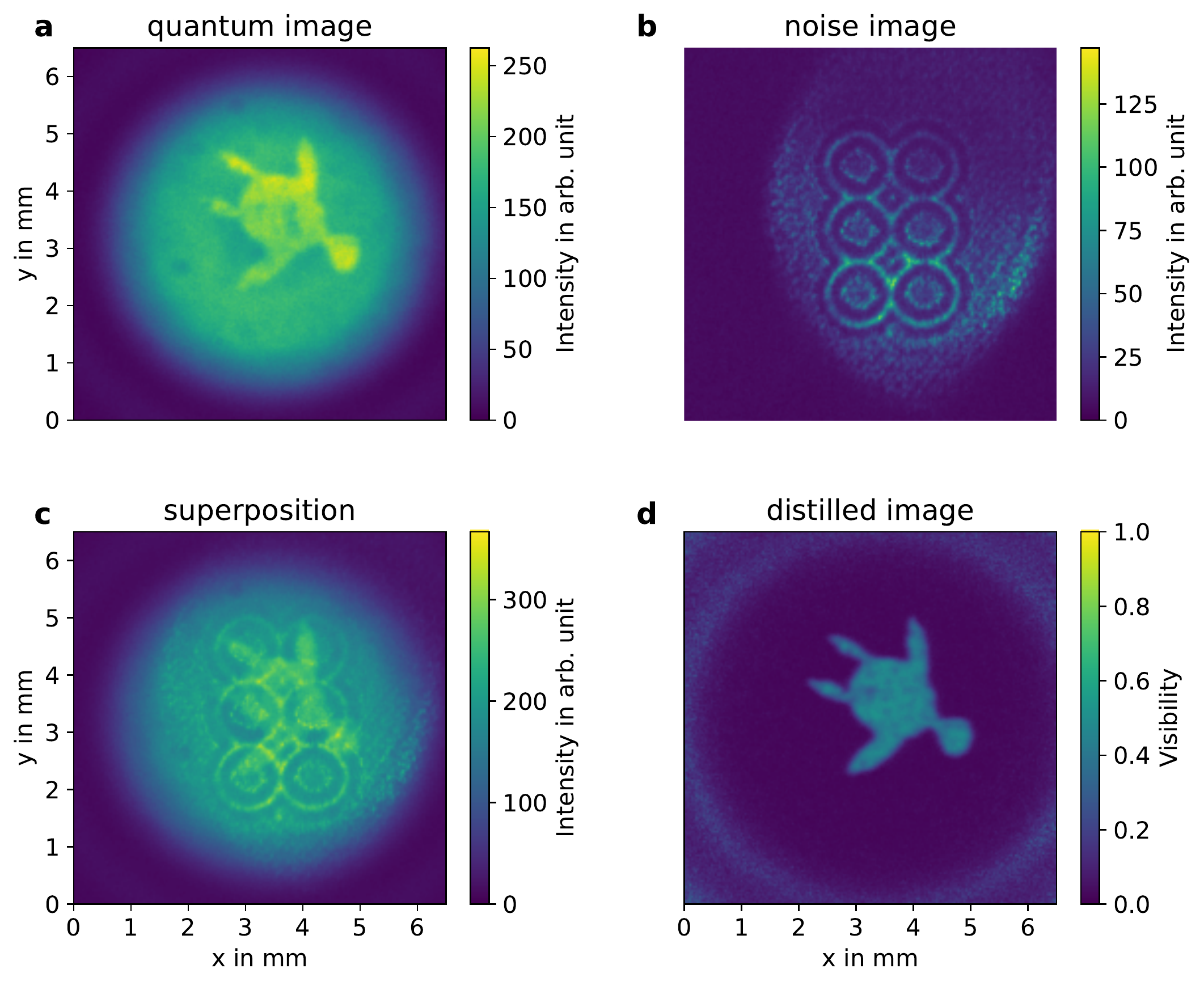}
	\caption{\textbf{Principle of quantum imaging distillation.} We employ a quantum imaging distillation protocol to remove a noisy image from a quantum image. (a) is the quantum image we aim to distill. (b) is the noise image that is superimposed on the quantum image. (c) is the superposition of noise and quantum images. (d) is the resulting distilled image from noise.} 
	\label{Fig:concept}
\end{figure}
Quantum imaging distillation is a process whereby a quantum image is cleaned from noise. To explain our distillation technique let us consider two images: a \textit{quantum image}, which is acquired by illuminating the sample with non-classical light, and a \textit{noise image}, which is an image detected at the camera and generated with classical illumination. These two images are shown in Figs.~\ref{Fig:concept}(a) and \ref{Fig:concept}(b), respectively. A noise image is an unwanted signal that is superimposed over a quantum image on the camera. This image superposition is shown in Fig.~\ref{Fig:concept}(c). To distill an image, different photon pair degrees of freedom can be employed, e.g., frequency, time, or spatial correlations. We use the amplitude modulation of quantum holography with undetected light (QHUL)~\cite{topfer2022quantum}, which is an interferometric quantum imaging technique~\cite{lemos:2014:quantum}. In QHUL, the object information is carried into a single-photon interference pattern. When the noise reaches the camera, if the intensity difference of QHUL is bigger than the intensity variance of the noise, the quantum image can be distilled. The resulting distilled image is shown in Fig.~\ref{Fig:concept}(d).     

\subsection*{Theory}

Spontaneous parametric down-conversion (SPDC) \cite{klyshko1969scattering,burnham1970observation} is a well-known nonlinear process that generates photon pairs (signal and idler) mediated by the interaction of an intense pump beam with the atoms of a nonlinear crystal~\cite{walborn2010spatial}. Our imaging scheme makes use on a SU(1,1) interferometer, wherein a pair of signal-idler photons can be generated in one of two propagation modes, forward and backward. The probability to generate paired photons in both modes (forward or backward) simultaneously is negligible~\cite{lahiri2019nonclassicality,wiseman2000induced}. In the forward propagation mode, the pump, signal, and idler beams are spatially separated, and later, back-reflected into the nonlinear crystal with the help of 4f systems. Before back reflection, the idler photon is reflected from an object, with complex reflectivity $R=|R|\exp(i\varphi_R)$, placed in front of its end-mirror. In the backward propagation mode, the idler photon does not interact with the sample. The signal photons are collected by a camera, and the idler photon remains undetected. 

\vspace{5mm}
The mean value of signal photons detected in time $T_D$ at one pixel of the camera of area $S_D$ (see supplementary material for details) is
\begin{align} \label{Eq:Sg-count-rate}
    \avgp{N_S}_{\delta}=2\:S_0\:[1+|R|\:\gamma\, \cos(\delta + \varphi_R)].
\end{align}
$\delta$ is an interferometric spatially invariant phase and $S_0$ is the number of signal photons generated in single-pass SPDC (in a time window $T_D$ and area $S_D$). The parameter $\gamma$ is related to the effective bandwidth of the signal-idler photon pairs, determined essentially by the bandwidth of the filters located in front of the camera \cite{topfer2022quantum}. From Eq.~\eqref{Eq:Sg-count-rate}, we see that $\avgp{N_S}$ changes when $\delta$ is varied. In particular, using $\delta=0, \pi/2, \pi$, and $3\pi/2$, the object's information can be retrieved by means of QHUL as follows:
\begin{align} \label{Eq:OBJ-QHUL}
   &|R|=2\times\frac{\left(\left[ \avgp{N_S}_{3\pi/2}-\avgp{N_S}_{\pi/2}\right]^2+\Big[\avgp{N_S}_{0}-\avgp{N_S}_{\pi}\Big]^2\right)^{1/2}}{\avgp{N_S}_{0}+\avgp{N_S}_{\pi/2}+\avgp{N_S}_{\pi}+\avgp{N_S}_{3\pi/2}},\\
     &\varphi_R=\arctan\left(  \frac{\avgp{N_S}_{3\pi/2}-\avgp{N_S}_{\pi/2}}{\avgp{N_S}_{0}-\avgp{N_S}_{\pi}} \right) .\label{Eq:OBJ-QHUL2}
\end{align}
Equations~\eqref{Eq:OBJ-QHUL} and \eqref{Eq:OBJ-QHUL2} are not unique representations of $|R|$ and $\varphi_R$ and, in general, these quantities can be extracted using different number of phases~\cite{topfer2022quantum}. We emphasize that in this technique paired photon coincidences are not needed and only signal photons are measured.  

\vspace{5mm}
For the important case of phase estimation, we evaluate the sensitivity of QHUL obtaining the variance of $\varphi_R$ given by Eq. \eqref{Eq:OBJ-QHUL2}. We first calculate the variance of the signal-photon flux $\avgp{(\Delta N_S)^2}=\avgp{ N_S^2}-\avgp{N_S}^2$. Since the coherence time $T_C$ of signal-idler photon pairs ($T_C \sim 1/B$, $B$ is the effective bandwidth of SPDC) is much smaller than the detection time, we can approximate the variance of the signal-photon flux to as (see Supplementary Material), 
\begin{equation}
\label{Eq:Sg-variance}
\avgp{( \Delta N_S )^2}=\avgp{N_S},
\end{equation}
 which is equivalent to considering Poissonian statistics. This result is well-known to be valid when considering a multimode signal where each mode has the same non-Poissonian statistics.

\vspace{5mm}
The result in Eq.~\eqref{Eq:Sg-variance} corresponds to the minimum signal variance achievable in QHUL. However, this variance can rapidly increase by electronic noise, e.g., camera signal-to-noise ratio, or external sources of noise such as temperature fluctuations, airflow, and external illumination, among others. In this work, we study the effect of an external classical illumination impinging on the camera, overlapping the quantum image of interest. Since QHUL detects only the single-photon stream of signal photons, the decoherence produced by an external source of light seems extremely harmful and, therefore, the image distillation highly improbable. However, we will demonstrate experimentally that the quantum imaging distillation is possible even in scenarios with  considerable high levels of noise in comparison to the quantum signal intensity.
\par
\par
\begin{figure}[ht]
	\centering	 \includegraphics[width=1\linewidth]{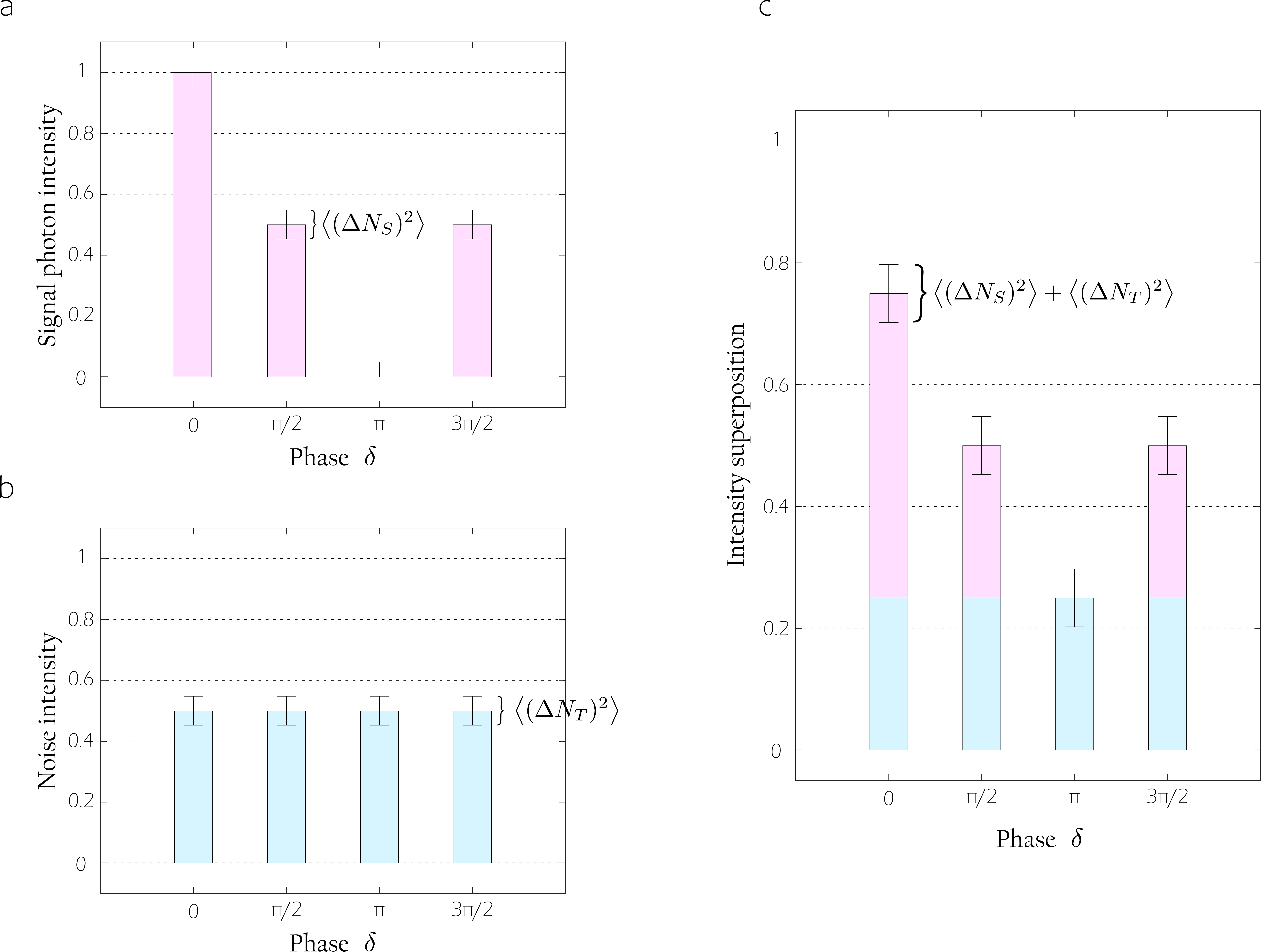}
	\caption{\textbf{Intensity detected in one camera pixel.} For visualization purposes, we have considered the same mean intensities (chart bars) and variances (error bars) for the signal photon and the noise. In (a) is presented the signal photon intensity (in pink) for different values of $\delta$. This intensity fluctuation allows us to compute the object information using QHUL~\cite{topfer2022quantum}. In (b) is presented the noise intensity (in blue) for the same phases $\delta$. In contrast to the signal intensity, the noise intensity is not affected by the value of $\delta$. In (c) are added the noise and signal intensities. Since noise intensity does not change, its contribution just sets a higher background. On top of it, signal photon intensity still changes, and the total variance is its previous variance plus the noise variance. Thus, an external source of noise increases the shot noise of QHUL.}
	\label{fig:distillation_int}
\end{figure}
\vspace{5mm}
Quantum holography with undetected light employs interferometric modulation of the signal photon to retrieve the object information, as shown in Eqs.~\eqref{Eq:OBJ-QHUL} and \eqref{Eq:OBJ-QHUL2}. We show that this modulation can also be used for distillation purposes, which is depicted in Fig.~\ref{fig:distillation_int}. For each value of the interference phase $\delta$, the signal photon has a well-defined intensity and variance, given by Eqs.~\eqref{Eq:Sg-count-rate} and \eqref{Eq:Sg-variance}, respectively. In Fig.~\ref{fig:distillation_int}(a), signal intensity (variance) is represented with pink bar charts (error bars). In contrast, the intensity of a stochastic noise $\avgp{N_T}$ fluctuates randomly with a variance $\avgp{(\Delta N_T)^2}$. For the sake of simplicity, we consider (but are not restricted to) the case where the noise has the same total mean intensity and variance than the signal photon. While the mean noise intensity is $\avgp{N_T}$, the mean signal intensity corresponds to $\{\textup{min}(\avgp{N_S})+\textup{max}(\avgp{N_S} )\}/2$. In Fig.~\ref{fig:distillation_int}(b), the noise intensity (variance) is represented with blue bar charts (error bars). As a result of adding these two intensities, see Fig.~\ref{fig:distillation_int}(c), the \textit{background} increases up to the noise intensity, while the signal intensity varies on top of it. The noise variance contributes to the signal intensity variance, i.e., the shot noise increases. In this way, one can infer that if the difference of the signal intensity is higher than the noise variance, the quantum image can be distilled. Additionally, the shot noise of QHUL always increases if the noise intensity and/or its variance increases. More details are given in the Supplementary Material.

\vspace{5mm}
Let us analyze the performance of our distillation technique by considering a generalization of Eq.~\eqref{Eq:OBJ-QHUL}. For QHUL with $M$ phase steps, we have that
\begin{align}
    \varphi_R=-\tan^{-1}\left(\sum_j \langle N_S\rangle_j \sin \delta_j / \sum_j \langle N_S \rangle_j \cos \delta_j \right),
\end{align}
with $\delta_j=j\:2\pi/M$, and $j=0,1,...,M-1$. The phase variance, including noise, it reads
\begin{align}\label{Eq:phase-estimation-variance}
    \avgp{(\Delta \varphi_R)^2} =\sum_j \left( \frac{\partial \varphi_R}{\partial \avgp{N_S}_j} \right)^2 \left[ \avgp{(\Delta N_S)^2}_j +\avgp{(\Delta N_T)^2} \right],
\end{align}
where 
\begin{align}\label{Eq:Derivative}
   \left( \frac{\partial \varphi_R}{\partial \avgp{N_S}_j} \right)^2=\frac{1}{M^2 \, \gamma^2 \, |R|^2 \, S^2_0 } \sin^2(\varphi_R+ \delta_j).
\end{align}
Replacing Eqs. \eqref{Eq:Sg-variance} and \eqref{Eq:Derivative} into Eq. \eqref{Eq:phase-estimation-variance}, and considering $n$ measurements, we obtain that the variance of phase estimation is
\begin{align} \label{Eq:Sg-variance-final}
    \avgp{(\Delta \varphi_R)^2}= \frac{1}{n \,S_0}\, \frac{1}{M \, V^2} \left\{ 1+ \frac{\avgp{(\Delta N_T)^2}}{2 \, S_0}\right\}.
\end{align}
$S_0$ is the number of idler photons that illuminate the object whose phase we want to estimate, $\avgp{(\Delta N_T)^2}$ is the variance of the number of background photons that illuminate the detector, and $V=|R| \gamma$ is the visibility of the signal-photon flux interference pattern as a function of the phase $\delta$. The visibility is defined by $V= \{\textup{max}(\avgp{N_S})-\textup{min}(\avgp{N_S})\}/\{\textup{max}(\avgp{N_S})+\textup{min}(\avgp{N_S})\}$. Equation~\eqref{Eq:Sg-variance-final} contains two contributions to the variance of the phase: the first term comes from the quantum illumination, and the second term comes from the noise illumination. While the former can be shown to be well described by Poissonian statistics (see Supplementary Material), the latter depends on the specific characteristics of the noise illumination.


\subsection*{Quantum imaging}
\begin{figure}[ht]
	\centering
\includegraphics[width=1\linewidth]{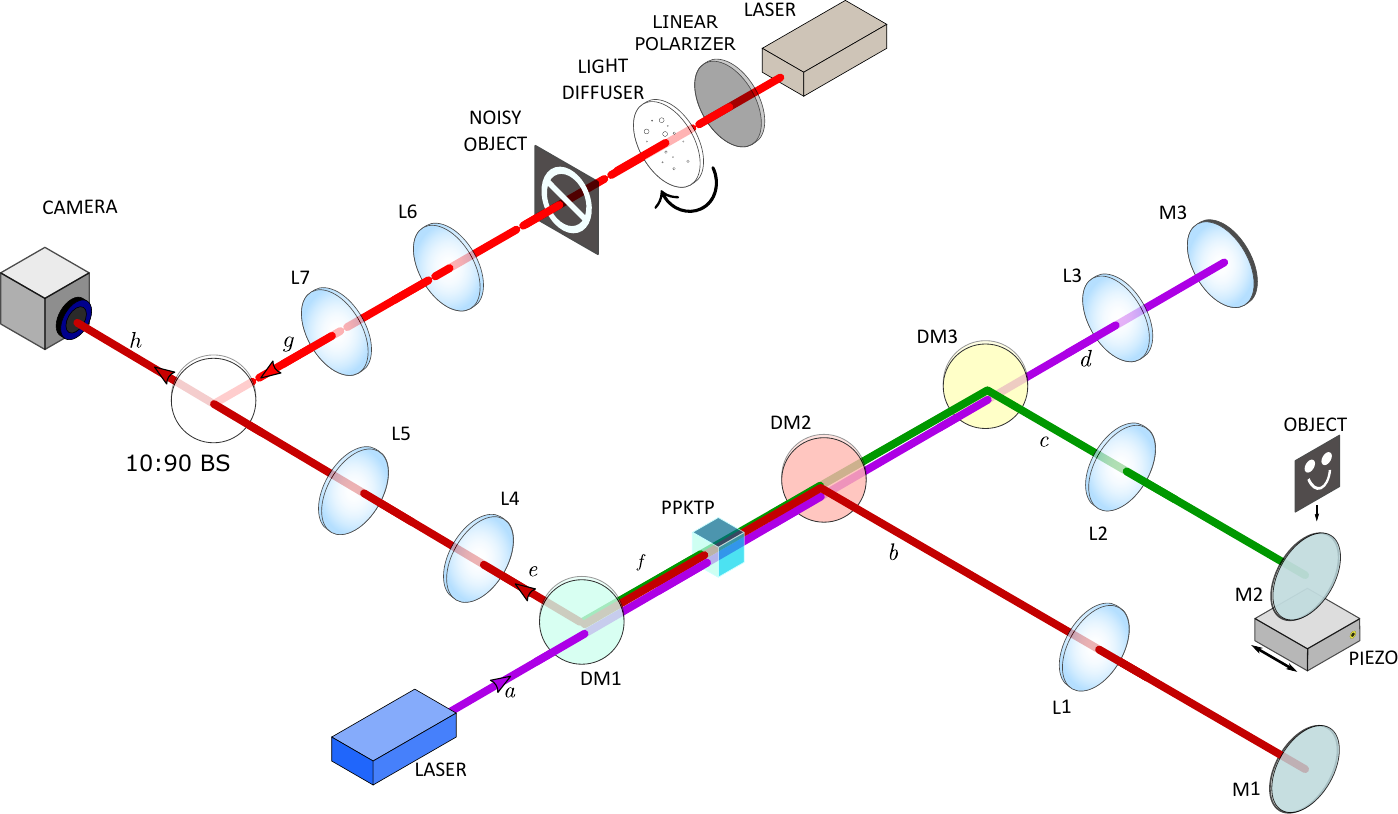}
	\caption{\textbf{Setup.} The signal and idler beams (in paths $b$ and $c$, respectively) are generated by the pump beam in path $a$ interacting with the nonlinear crystal (ppKTP) in the forward direction, while paths $e$ and $f$ represent propagations of the down converted beams generated after the pump beam is reflected back into the crystal by mirror M3 in path $d$. An object in path $c$ is illuminated with the idler beam in the Fourier plane of the crystal. To create the noise, a laser diode of the same wavelength as the signal photon (910 nm) is used. The signal beam in path $e$ is merged with the noise in path $g$ before reaching the camera with a 10:90 beam splitter. On the detector plane we obtain the quantum image with lenses L2, L4 and L5, and the noise image with lenses L6 and L7. Different type of noise are created with a linear polarizer and a light diffuser in path $g$. The linear polarizer controls the pump power of the diode laser. The diffuser that consists on a rotating ground glass plate produces a speckle pattern of the noise source. The speed of the rotation is controlled through the glass plate motor interface.}
	\label{fig:setup}
\end{figure}

A sketch of our experimental implementation is depicted in Fig.~\ref{fig:setup}. For our quantum image, we used a SU(1,1) nonlinear interferometer~\cite{yurke1986INTER} in a Michelson configuration, where its input/output is a nonlinear medium. Our crystal is a periodically poled potassium titanyl phosphate (ppKTP) of $2\times2\times1$~mm$^3$ (length $\times$ width $\times$ height), which is pumped bidirectionally ($a$ and $f$ directions) by a CW laser at 405~nm and with avarage power of 90 mW. Due to its strong $\chi^{(2)}$--nonlinearity, a photon pair (signal and idler photons), is generated through SPDC into the paths $a$ or $f$, but never simultaneously. Signal (idler) photons have a central wavelength of $\lambda_S=910$ nm ($\lambda_I=730$ nm). 

\vspace{5mm}
In the forward propagation direction $a$, signal, idler, and pump beams are spatially separated with dichroic mirrors DM2 and DM3 into the paths $b$, $c$, and $d$, and reflected back into the crystal with mirrors M1, M2, and M3.
In front of mirror M2, an object with a complex amplitude $R=|R|\exp(i\varphi_R)$ is placed. In $b$, $c$, and $d$, lenses of focal length f$=150$ mm transform transverse position (source plane) into transverse momentum (mirrors plane). Therefore, in path $c$, a wave vector $\kk_I$ representing a plane wave of the idler photon is focused to a point on the object~\cite{lahiri2015theory,fuenzalida2022resolution}. The interaction of idlers being absorbed or reflected by the object can be modeled with the help of a beam splitter~\cite{vega2022fundamental}.
In the backward propagation $f$, signals are collected by a camera and idlers remain undetected. We ensured that by placing a 800~nm long-pass filter and a $910\pm1.5$~nm interference filter in front of the camera. Our camera is the Prime BSI Scientific CMOS from Teledyne Photometrics with a pixel size of 6.5 $\mu$m. Signal photon's transverse momentum is obtained with the lens L4 of focal length f$=100$~mm performing a Fourier transform of the source plane. This plane is later imaged on camera with the lens L5 of focal length f$=150$~mm. Thus, a wave vector $\kk_S$ of the signal photon is focused to a point on the camera. If $a$ and $f$ propagation are perfectly aligned, the photon pair emission (which-source) information is erased. Consequently, on the camera, an interference pattern of signal photons is observed~\cite{zou:1991:induced}. Moreover, the object information obtained herein by the idler photon, is transferred to the signal photon interference pattern~\cite{lemos:2014:quantum,gilaberte2021video}, see Eq.~\eqref{Eq:Sg-count-rate}. The interferometric phase $\delta$ is changed with a piezo placed below mirror M2. The object information is retrieved by employing QHUL of 12 steps~\cite{topfer2022quantum} with an acquisition time of $T_D$=1~s per image.

\vspace{5mm}
\subsection*{Noise source}
A CW diode laser of $\lambda_N=910$~nm and variable pump power is employed to introduce noise in the system. The laser illuminates an object, which is imaged on the camera with a 4f system using the lenses L6 and L7 of focal lengths f$=150$~mm and f$=125$~mm, respectively. This classical image is superimposed on top of the quantum image on the camera using a 10:90 beam splitter, see Fig.~\ref{fig:setup}. Properties of classical illumination, intensity and variance, are changed in order to evaluate the effects of noise in QHUL and our distillation performance. Experimental details about the noise properties can be found in the Supplemental Material.

\vspace{5mm}
\subsection*{Distillation performance to different noise intensities}
\begin{figure}
	\centering
		\includegraphics[width=1\linewidth]{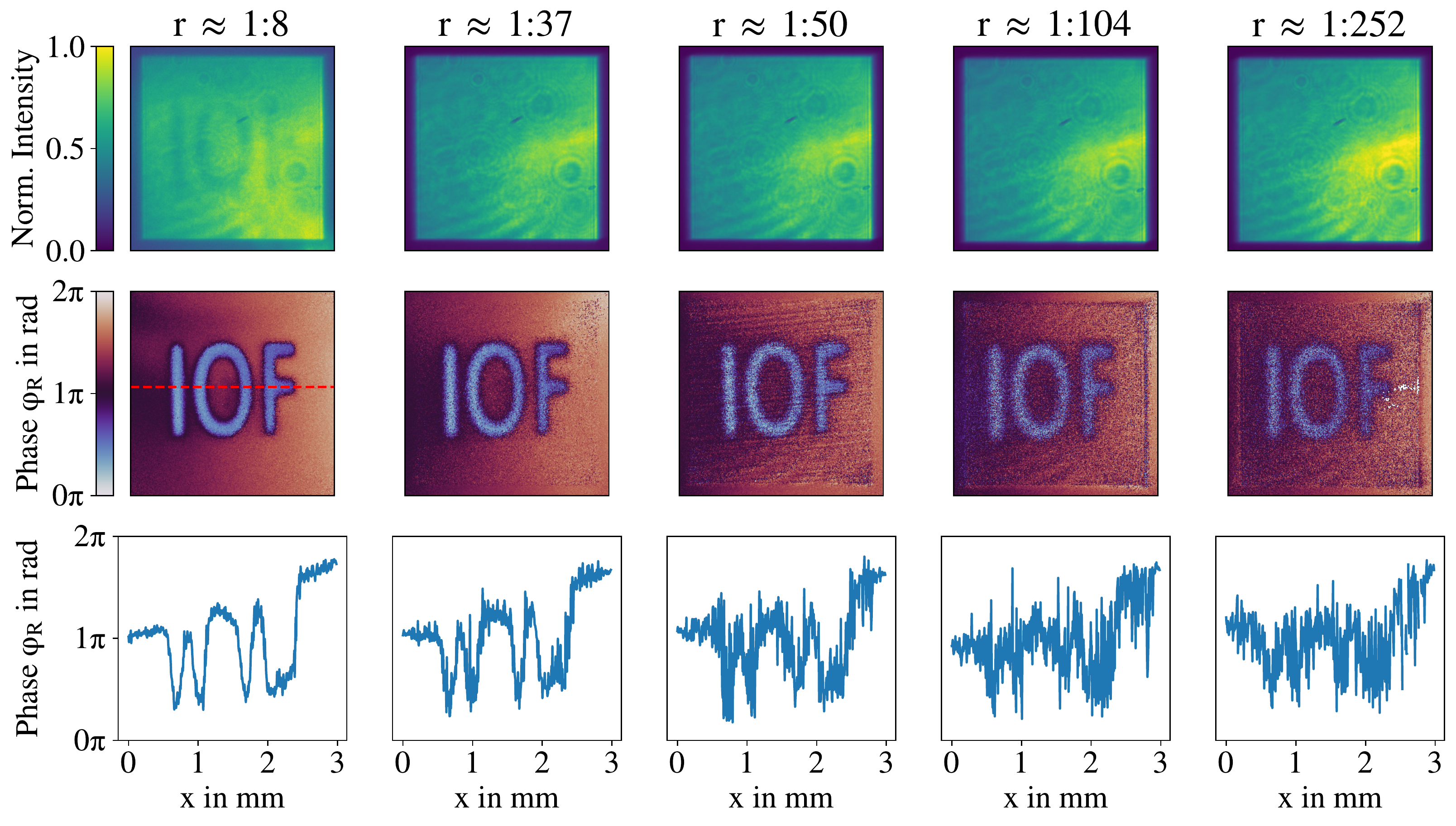}
	\caption{\textbf{Resilience to different noise intensities.} In the upper row are shown superpositions of quantum (IOF letters) and classical (square shape) images. The ratio between their mean intensities is stated on top of each image. In the middle row are presented the experimental results for our distillation technique through QHUL. In the last row are presented a transverse cut of the distilled images. We observe that while the noise intensity increases, the phase estimation diminishes.}
	\label{fig:noise_levels}
\end{figure}

In the first experiment, while having superimposed classical and quantum images, QHUL is performed to distill the quantum image under different intensities of noise. We first characterized the signal flux rate emission measuring its mean intensity of an illuminated area on the camera. For the quantum image $S_0=$~134 signal photons generated in a single pass through the crystal were employed. The detection window was $T_D=1~\rm{s}$ and the detection area was $S_D\approx 32.5 \times 32.5~\rm{\mu m}^2$. Signal intensity does not change during experiments. In a similar way but independently measured, different noise intensities are characterized, which are obtained by changing the angle of a linear polarizer (LP) in front of the laser in path $g$. The experiment starts superimposing the quantum and classical images on the camera.
Experimental results are shown in Fig.~\ref{fig:noise_levels}. Its first row shows the superposition of classical and quantum images; noise intensity increases from left to right with the following ratios ($r=$ mean signal intensity: mean noise intensity), $r\approx1:8$, $r\approx1:37$, $r\approx1:50$, $r\approx1:104$, and $r\approx1:252$. Second row in Fig.~\ref{fig:noise_levels} shows distilled images by QHUL of the corresponding upper superposed images. Imaging distillation through QHUL is successfully achieved in every case, even with a noise intensity 250 times higher than the signal intensity. However, we notice that while noise intensity increases, sharpness of distilled images decreases. One can observe this in detail in the third row that shows a transverse cut of the distilled images (represented by a dotted red line). It is clear from our experimental results that phase accuracy diminishes as the noise intensity increases, which also corresponds to prediction given in Eq.~\eqref{Eq:Sg-variance-final}.

\vspace{5mm}
\subsection*{Induced variance by noise}
\begin{figure}
	\centering
\includegraphics[width=.8\linewidth]{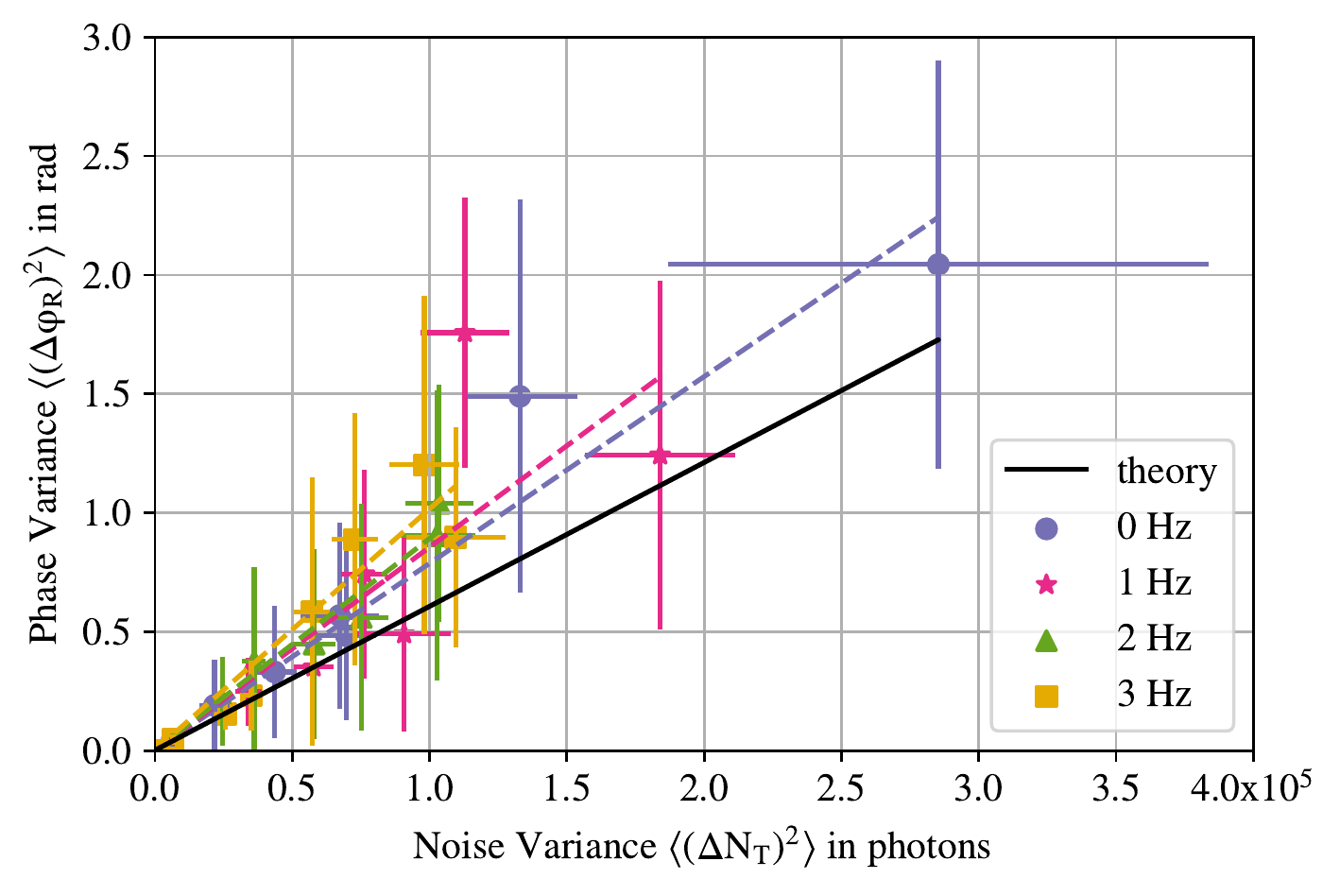}
	\caption{\textbf{Distillation phase variance affected by noise variance.} A light diffuser with four different rotation speeds is employed to change the properties of the noise illumination, see App.~D. The different noise configurations are represented by different colors and symbols, see inset. Data points represent the experimental phase values obtained for different noise variances and dashed lines represent their fits. A theoretical black solid line representing a Poissonian noise is also included. In all configurations we observed that a higher noise variance increases the phase inaccuracy in QHUL. We also corroborate that the phase sensitivity is linearly dependent with the noise variance, more details in App.~F.}
	\label{fig:speed}
\end{figure}

In a second experiment, we quantified the effects of noise variances on the phase accuracy of distilled images. The same configurations of noise intensities are employed. Additionally, a light diffuser mounted on a rotational motor with four angular frequencies of 0, 1, 2, and 3 Hz changed the noise variance. The noise variance is characterized considering the intensity variation of one pixel over 12 consecutive images. Experimental results are shown in Fig.~\ref{fig:speed}. We plot the experimental measured values for the phase variance of distilled images against the noise variance for different angular frequencies (data point with error bars; purple circle = 0 Hz; rose star = 1 Hz; green triangle = 2 Hz; yellow square = 3 Hz). We also provide a theoretical prediction for comparison (solid black line). 

The results show that an increment of the noise variance increases proportional with the phase variance. Also it can be observed that in every case a linear dependence appears between these two variances, see App. F for more details. However, the noise variance introduces a slightly higher phase variance than expected. A reasonable explanation for this is that additional sources of noise were involved during the measurement process, such as airflow or temperature fluctuations. The experimental behaviour of variances is in good agreement to theoretical predictions presented above in Eq.~\eqref{Eq:Sg-variance-final}. 
\vspace{5mm}
\par
To conclude, we compare our distillation to other previously introduced techniques in Tab.~1. In our implementation, we have used at least five times more noise than in all previous related experiments, showing the highest resilience to date.

\begin{table}[H]
\centering
\begin{tabular}{||c|c||}
\hline
Distillation techniques & Signal-to-noise ratio \\
\hline
Phys. Rev. A \cite{duncan2019quantum} &  1\,:\,0.14 \\
\hline
Sci. Adv. \cite{defienne2019quantum} & 1\,:\,10 \\
\hline
Phys. Rev. A \cite{zhang2020multidimensional} & 1\,:\,49 \\
\hline
Sci. Adv. \cite{gregory2020imaging}  & 1\,:\,5.8 \\ 
\hline
Sci. Rep. \cite{gregory2021noise} & 1\,:\,20\\
\hline
Our work & 1\,:\,252\\
\hline
    \end{tabular}
    \caption{\textbf{Distillation performances.}}
    \label{tab:dist.works}
\end{table}

\section*{Discussion}
Our work explores for the first time the effects of noise in quantum imaging with undetected light. We have also introduced a technique to distill the quantum image from that noise. Our quantum imaging distillation technique is based on quantum holography with undetected light~\cite{topfer2022quantum}. This technique uses a photon pair, signal and idler, where idler illuminates the object and signal is detected on the camera. The idler photon remains undetected and its phase modulation is employed in the imaging distillation procedure.
\vspace{5mm}

To prove our technique, we superimposed partially or completely a classical source of noise on top of our quantum image on the camera. Our technique worked in every occasion, even for noise intensities 250 times higher than our signal intensity. However, the noise variance does affect the phase accuracy of our distillation technique. A higher noise variance produces a higher phase inaccuracy, where in general, these two quantities are linearly dependent. 
\vspace{5mm}

\begin{figure}[htpb]
	\centering
\includegraphics[width=1\linewidth]{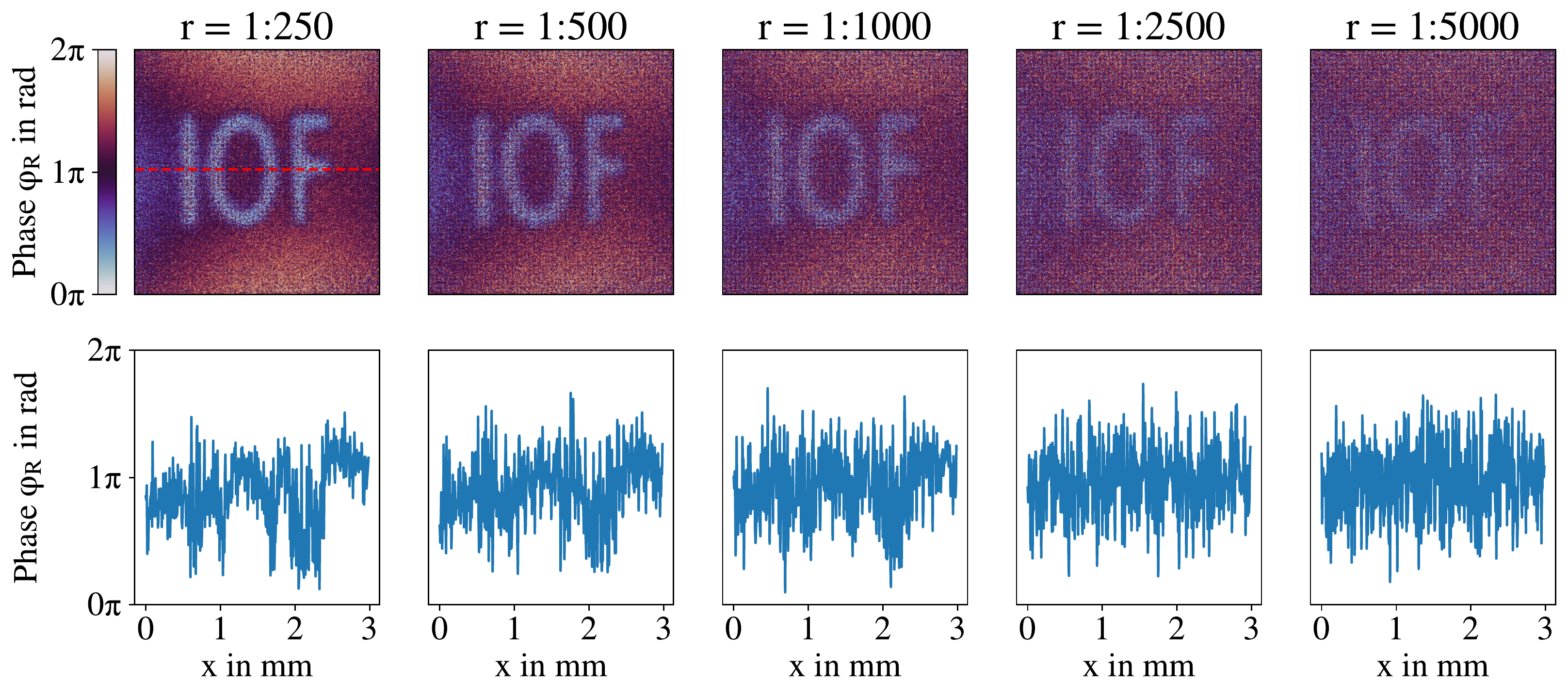}
\caption{\textbf{Simulated resilience limits.} In order to find the limits of our technique, we have simulated a Poisonnian source of noise superimposed on our quantum image. The first row shows distilled images for different ratios stated above them. Second row shows a transverse cut of distilled images on top. The simulations show that our technique should be able to work at noise levels beyond 1000 times the quantum signal.}
\label{fig:simulations}
\end{figure}

To extend our work and explore the limits of our technique, in Fig.~\ref{fig:simulations}, we present simulations of QHUL under extreme noise scenarios. For this, we have considered a noise with Poissonian statistics and with ratios up to 5000 times higher than the mean intensity of the signal of interest. The simulations show that our distillation technique keeps working until $r=$1000, or even until $r=2500$. For $r=5000$, the distilled image is already blurred, however, it still exhibits some features of the original object.

\vspace{5mm}
Although, in our experiment we employed a classical source of noise, this distillation technique should also work for a quantum source of noise. For example, by replacing the noise laser with a phase-matched SPDC source and performing far field or near field imaging of the noisy object on the camera. Furthermore, since the underlying principle of this distillation technique is the phase modulation, our technique should be applicable to QIUL based on position correlations~\cite{viswanathan2021resolution,kviatkovsky2022mid}. 
\vspace{5mm}

Our results are not just vital for QIUL but can have an important contribution to other techniques based on induced coherence without induced emission~\cite{hochrainer2022review}, such as spectroscopy~\cite{kalashnikov2016infrared}, sensing~\cite{kutas2020terahertz}, optical coherence tomography~\cite{valles2018optical,paterova2018tunable}, entanglement certification~\cite{lemos2023one}, and quantum state tomography~\cite{fuenzalida2022quantum}.
\vspace{5mm}

Our experiment is a step forward for quantum imaging in open systems and could be useful to the understand the limitations of a (quantum) LIDAR with undetected light.
\section{Acknowledgments}
The authors thank Patrick Hendra and Fabian Steinlechner for fruitful discussions. This work was supported as a Fraunhofer LIGHTHOUSE PROJECT (QUILT). Furthermore, funding support is acknowledged from the German Federal Ministry of Education and Research (BMBF) within the funding program Photonics Research Germany with contract number 13N15088. In addition, funding is acknowledged from the Thüringer Aufbaubank via the projects Multi-Use (2020FGI0023), SPEQTRA (2021FE9016), and Quantum Hub Thuringia (2021FGI0041). J.P.T. acknowledges the funding by the R\&D project CEX2019-000910-S, funded by the Ministry of Science and innovation (MCIN/ AEI/10.13039/501100011033/), from Fundació Cellex, Fundació Mir-Puig, and from Generalitat de Catalunya through the CERCA program. J.P.T. also acknowledges support from the project 20FUN02 “POLight”, that has received funding from the EMPIR programme co-financed  by the  Participating  States  and  from  the  European  Union's  Horizon  2020  research  and innovation programme, and financial support from project QUISPAMOL (PID2020-112670GB-I00) funded by MCIN/AEI /10.13039/501100011033.

\section{References}

\bibliography{biblio}

\end{document}


\title{Supplementary Material of ``Experimental quantum imaging distillation with undetected light''}

\author{Jorge Fuenzalida,$^{1,2,\dagger,\star,}$  Marta Gilaberte Basset,$^{1,3,\star}$ Sebastian T{\"o}pfer,$^{1,2,\star}$ Juan P. Torres,$^{4,\ddagger}$ and Markus Gr{\"a}fe$^{1,2,3,\S}$
\vspace{1em}
\\
\it $^{1}$Fraunhofer Institute for Applied Optics and Precision Engineering IOF, Albert-Einstein-Str. 7, 07745 Jena, Germany. \\ \vspace{0.5em}
\it $^{2}$Technical University of Darmstadt, Department of Physics, Institute of Applied Physics, Schloßgartenstraße 7, 64289 Darmstadt, Germany.\\ 
\vspace{0.5em}
\it $^{3}$Friedrich-Schiller-University Jena, Abbe Center of Photonics,\\ 
\it Albert-Einstein-Str. 6, 07745 Jena, Germany.\\ \vspace{0.5em}
\it $^{4}$ICFO-Institut de Ciencies Fotoniques, The Barcelona Institute of Science and Technology,\\
\it 08860 Castelldefels, Spain, and Dept. Signal Theory and Communications, Universitat Politecnica de Catalunya, 08034 Barcelona, Spain.\\
\vspace{0.5em}
$^{\dagger}$ jorge.fuenzalida@tu-darmstadt.de\\
$^{\ddagger}$ juanp.torres@icfo.eu\\
$^{\S}$ markus.graefe@tu-darmstadt.de\\
$^{\star}$These authors contributed equally.
}

\date{\today}

\maketitle

\section{A. The quantum state of signal photons.}
In the undepleted pump beam approximation, if the bandwidth of parametric down-conversion (in the frequency and transverse wavenumber domains) is much larger than the corresponding bandwidth of the pump beam, the relationship between the quantum operators $a_S(\Omega,{\bf q})$ and $a_I(\Omega,{\bf q})$ (output face of the nonlinear crystal) and the operators $b_S(\Omega,{\bf q})$ and $b_I(\Omega,{\bf q})$ (input face of the nonlinear crystal) can be written in the low parametric gain approximation as \cite{dayan2007}
\begin{align}
\label{dayan2007}
a_S(\Omega,{\bf q})=& b_{S}(\Omega,{\bf q})  \exp \Big[ i k_{S}(\Omega,{\bf q})L \Big]  \nonumber \\ & + F_S(\Omega,{\bf q})\int \d \omega_P\, \d {\bf q}_P\, E_P(\Omega_P,{\bf q}_P) \,b_I^{\dagger}(\Omega_P-\Omega,{\bf q}_P-{\bf q}), \nonumber \\
a_I(\Omega,{\bf q})=& b_I(\Omega,{\bf q})  \exp \Big[ i k_{I}(\Omega,{\bf q}) L \Big] \nonumber \\ & + F_I(\Omega,{\bf q}) \int \d \omega_P\, \d {\bf q}_P\, E_P(\Omega_P,{\bf q}_P) \,b_{S}^{\dagger}(\Omega_P-\Omega,{\bf q}_P-{\bf q}),
\end{align}
where
\begin{equation}
 F_{S,I}(\Omega,{\bf q})= -i (\beta L)\, \text{sinc} \left[ \frac{\Delta_{S,I}(\Omega,{\bf q}) L}{2} \right]   \exp \left\{ i \left[k_P^0+ k_{S,I}(\Omega,{\bf q})-k_{I,S}(-\Omega,-{\bf q}) \right] \frac{L}{2} \right\} . \label{V_cw} 
\end{equation}
The nonlinear coefficient $\beta$ is 
\begin{equation}
\beta=\left( \frac{\hbar \, \omega_P \, \omega_S \,  \omega_I \, [\chi^{(2)}]^2}{64 \, \pi^3 \, \epsilon_0 \, c^3 \, n_P \, n_S \, n_I} \right)^{1/2},
\end{equation}
$n_{S,I,P}$ are refractive indexes, $\chi^{(2)}$ is the second-order nonlinear coefficient of the crystal, $L$ is the crystal length, and $\omega_{P,S,I}$ are central frequencies. The phase mismatch functions are $\Delta_{S,I}=k_P^0-k_{S,I}(\Omega,{\bf q})-k_{I,S}(-\Omega,-{\bf q})$, $k_{S,I}$ are the wavenumbers of signal and idler waves, and $k_P^0$ is the wavenumber of the pump beam. The total number of photons $N_P$ carried by the pump beam is
\begin{equation}
N_P=\int \d \Omega\, \d {\bf q}\, \big| E_P(\Omega, {\bf q}) \big|^2.
\end{equation}

\vspace{5mm}
If we consider that signal and idler photons $S_1$ and $I_1$ (forward propagation modes) are re-injected back into the nonlinear crystal with the help of 4f systems, the operator associated to signal photon $S_2$ (backward propagation mode) is
\begin{eqnarray}
& & a_{S_2}(\Omega,{\bf q})= F_S(\Omega,{\bf q})\, \Big\{\exp \big[ i k_{S}(\Omega,{\bf q}) L + i \varphi_S(\Omega) +i\varphi_{P_1}\big] \nonumber \\
& & +R^* \exp \big[ -i k_{I}(-\Omega,-{\bf q}) L + i\varphi_{P_2}-i \varphi_I(-\Omega) \big] \Big\}  \int \d \omega_P\, \d {\bf q}_P\, E_P(\Omega_P,{\bf q}_P) \,b_I^{\dagger}(\Omega_P-\Omega,{\bf q}_P-{\bf q}) \label{signal2} \nonumber \\
& & + F_S(\Omega,{\bf q})\,\exp \big[ i\varphi_{P_2}\big]\, \int \d\Omega_P\, \d{\bf q}_P\, E_P(\Omega_P,{\bf q}_P)\, f^{\dagger}(\Omega_P-\Omega,{\bf q}_P-{\bf q})  .
\end{eqnarray}
Notice that in Eq.~\eqref{signal2}, for the sake of simplicity, we have omitted the terms that depend on the operator $b_S$ since they will yield a zero contribution to the flux density of signal $S_2$ photons. We have made use of the approximation $k_{I}(\Omega_P-\Omega,{\bf q}_P-{\bf q})\sim k_{I}(-\Omega,-{\bf q})$, since the bandwidth of the pump beam is assumed to be very small when compared with the bandwidth of parametric down-conversion. We also consider that the phases of the pump beams, $\varphi_{P_1}$ and $\varphi_{P_2}$, that illuminate the nonlinear crystal might be different. The phases $\varphi_S(\Omega)$ and $\varphi_I(\Omega)$ are phases acquired by the signal $S_1$ and idler $I_1$ photons that traverse  4f systems before being re-injected back into the nonlinear crystal. The reflection coefficient of the object located in the idler $I_1$ path is $R=|R|\exp(i\varphi_R)$. The operator $f$ takes into account~\cite{RWboyd2008} 
the presence of the object with reflectivity $R$. These operators fulfill the relationships $\Big[f(\Omega,{\bf q}),f^{\dagger}(\Omega^{\prime},{\bf q}^{\prime}) \Big]=\big(1-|R|^2 \big)\, \delta(\Omega-\Omega^{\prime})\, \delta({\bf q}-{\bf q}^{\prime})$.

\section*{B. Flux density of signal photons detected in a 2f system}
We detect the flux density of signal photons $S_2$ with the help of a 2f system with focal length $f$. The input-output relationship between operators is similar to the classical relationship~\cite{Goodman2005} 
\begin{equation}
\label{eq:a2fsystem}
    a_{S_2}(t,{\bf x}_2) = \frac{1}{\lambda_S\,f}\,  \int \d{\bf x}_1\, a_{S_2}(t,{\bf x}_1) \exp \left[ -i\frac{2\pi}{\lambda_S\,f}\,{\bf x}_1 \cdot {\bf x}_2 \right],
\end{equation}
where ${\bf x}_1$ and ${\bf x}_2$ are the corresponding transverse coordinates in the input and output planes of the 2f system. For the sake of simplicity, we ignore any global phase. If we introduce the Fourier transform of the operator $a_{S_2}(t,{\bf x}_1)$ as $a_{S_2}(t,{\bf x})=(2\pi)^{-3/2}\, \int \d\Omega\, \d{\bf q}\, a_{S_2}(\Omega,{\bf q})\, \exp \Big[ i {\bf q} \cdot {\bf x}-i\Omega t \Big]$, we can write Eq.~\eqref{eq:a2fsystem} as
\begin{equation}
\label{eq:a2fsystem2}
    a_{S_2}(t,{\bf x}_2)=\frac{(2\pi)^{1/2}}{\lambda_S\,f}\, \int \d \Omega\,  a_{S_2}\left(\Omega,\frac{2\pi}{\lambda_S\,f} \, \text{\textbf{x}}_2\right)\, \exp \left( -i\,\Omega \,t \right),
\end{equation}
where we have made use of the identity $\int \d{\bf x}\,  \exp \Big[ i \big( {\bf q} - {\bf q}^{\prime} \big) \cdot {\bf x} \Big] = (2 \pi)^2 \delta({\bf q} - {\bf q}^{\prime})$. The flux density (photons/$m^2$) of signal photons detected at position ${\bf x}_2$  is 
\begin{equation}
\label{Nx}
N({\bf x}_2)=\frac{(2\pi)^2}{(\lambda_S f)^2} \int \d\Omega\,\left\langle a_S^{\dagger}\left(\Omega,\frac{2\pi}{\lambda_S f}\,{\bf x}_2\right)\,a_S \left(\Omega,\frac{2\pi}{\lambda_S f}\,{\bf x}_2\right)\right\rangle.
\end{equation}
If we make use of Eq.~\eqref{signal2} into Eq.~\eqref{Nx} we obtain
\begin{eqnarray}
& & N({\bf x}_2)=\frac{(2\pi)^2}{(\lambda_S f)^2}\,N_p \int \d\Omega\, \left|F_S\left(\Omega,\frac{2\pi}{\lambda_S f}\,{\bf x}_2\right) \right|^2 \Big\{  1-|R|^2 \nonumber \\
& & +\Big| \exp \big[ i k_{S}(\Omega,{\bf q}) L+i \varphi_S(\Omega) \big] +R^*\, \exp \big[-i k_{I}(-\Omega,-{\bf q}) L -i \varphi_I(-\Omega) \big] \Big|^2 \Big\} , 
\end{eqnarray}
where $N_P=\mathcal{I}_P S_P T_P$. $\mathcal{I}_P$ is the peak intensity of the pump beam, $S_P$ its area, and $T_P$ its time duration. If we make the substitution $\beta^2=\sigma/[(2\pi)^{3} \mathcal{I}_P]$, define  $V_S(\Omega) \equiv V_S(\Omega,{\bf x}_2=0)$, and associate a large value of $T_P$ with the detection time $T_D$, the number of photons detected in a small area $S_D$ centered around ${\bf x}_2=0$ is
\begin{equation}
N_0=\frac{S_P S_D}{(\lambda_S f)^2}\,\frac{T_D}{\pi} \int \d\Omega\, \Big| V_S(\Omega) \Big|^2 \Big\{  1+|R|\cos(\theta+\varphi_R) \Big\} .
\label{final}
\end{equation}
$V_S$ is the same functions as $F_S$ in Eq. (\ref{V_cw}) by substituting the nonlinear coefficient $\beta$ by $\sigma$. 
We should notice that similar expressions to Eq.~\eqref{final} for describing flux rates with detection systems based on 2f systems has been derived  using other methods \cite{Brambilla2001,Brambilla2004}.

\vspace{5mm}
If we approximate the phase matching function as $\Delta_S=D \Omega$, where $D$ is the difference of inverse group velocities at the central frequencies between signal and idler photons,
\begin{equation}
\theta=k_P^0 L+\frac{\omega_S L_S+\omega_I L_I}{c}+ \Omega \left[DL+\frac{L_S-L_I}{c}\right]+\varphi_{P_1}-\varphi_{P_2}.
\end{equation}
$L_{S,I}$ are the lengths of the 4f systems traversed by signal and idler photons, respectively, before being re-injected back into the nonlinear crystal. For the sake of simplicity, we write 
\begin{equation}
    |V(\Omega)|^2=\exp\left( -\pi \, \frac{\Omega^2}{B^2} \right).
\end{equation}
where $B$ is the bandwidth of parametric down-conversion. Making the integration over frequency in Eq.~\eqref{final}, we obtain
\begin{equation}
\langle N_0 \rangle=2S_0\,  \left[ 1+|R|\gamma \cos (\delta +\varphi_R) \right],
\end{equation}
with $\delta=k_P^0 L+(\omega_S L_S+\omega_I L_I)/c$,
\begin{equation}
S_0=\frac{S_P \, S_D}{(\lambda_S f)^2} \frac{T_D B}{2\pi}\,(\sigma L)^2,
\end{equation}
and
\begin{equation}
\gamma=\exp \left[ -\frac{B^2}{4\pi}\, \left( DL+\frac{L_s-L_i}{c}\right)^2\frac{}{}\right].
\end{equation}
Notice that the visibility of signal $\langle N_0 \rangle$ as function of phase $\delta$ is 
\begin{equation}
\label{visibility}
V=|R|\gamma.
\end{equation}
$N_0$ is the variable that we designate as $N_S$ in the main text.

\section*{C. Sensitivity of phase estimation with phase shifting digital holography under the presence of external noise}
For the sake of simplicity, we write below $N_S=N_0 \equiv N$. We aim at estimating $\varphi_R$ using $M$ phases ($\delta_j=j \,2\pi/M$ with $j=0,1,...,M-1$). The expression for the estimation  of the phase is
\begin{equation}
\varphi_R=-\tan^{-1}\,\frac{\sum_j \avgp{N}_j \sin (\delta_j)}{\sum_j \avgp{N}_j \cos (\delta_j)}  .
\end{equation}
For M=4 (phases $\delta_j=0,\pi/2,\pi,3\pi/2$) we have
\begin{equation}
\varphi_R=\tan^{-1}\,\frac{\avgp{N}_{3\pi/2}-\avgp{N}_{\pi/2} }{\avgp{N}_0-\avgp{N}_{\pi}}  .
\end{equation}
We will make use of the error propagation formula
\begin{equation}
\label{errorformula1}
\langle (\Delta \varphi_R)^2\rangle=\sum_j \left( \frac{\partial \varphi_R}{\partial \avgp{N}_j}\right)^2 \Big[ \langle (\Delta N)^2\rangle_j +\avgp{(\Delta N_T)^2} \Big]. 
\end{equation}
where $\avgp{N_T}$ is the background signal (independent of the signal of interest) that reaches the detector with a variance of $\avgp{(\Delta N_T)^2}$. If the signal $N_j$ has a bandwidth $B$, and the detection time $T_D$ is large, i.e., $T_D \gg 1/B$, a condition that applies in the experiment, we can safely write that $\langle (\Delta N)^2\rangle_j=\langle N\rangle_j$. Therefore, we can write
\begin{equation}
\label{errorformula2}
\langle (\Delta \varphi_R)^2\rangle=\sum_j \left( \frac{\partial \varphi_R}{\partial \avgp{N}_j}\right)^2 \Big[ \langle N \rangle_j +\avgp{(\Delta N_T)^2} \Big]  .
\end{equation}

\vspace{5mm}
The derivatives are
\begin{equation}
\frac{\partial \varphi_R}{\partial \avgp{N}_i}= \frac{\sum_j \avgp{N}_j \sin \delta_j \cos \delta_i-\sum_j \avgp{N}_j \cos \delta_j \sin \delta_i }{\left[ \sum_j \avgp{N}_j \sin \delta_j \right]^2+ \left[ \sum_j \avgp{N}_j \cos \delta_j \right]^2} .
\end{equation}
After some calculations, we obtain
\begin{eqnarray}
& & \sum_j \avgp{N}_j \sin (\delta_j)=-M|R| \gamma\, S_0 \sin (\varphi_R) ,\nonumber \\
& & \sum_j \avgp{N}_j \cos (\delta_j)=M|R| \gamma\,S_0  \cos (\varphi_R) ,
\end{eqnarray}
where we have made use of
\begin{equation}
\sum_i \sin^2(\varphi_R+\delta_j) =\frac{M}{2},
\end{equation}
and
\begin{equation}
\sum_j \sin^2(\varphi_R-\delta_j) \cos (\varphi_R+\delta_j)=0.
\end{equation}
We can easily verify numerically the validity of these expressions for several values of $M$. Finally, we obtain 
\begin{equation}
\left( \frac{\partial \varphi_R}{\partial \avgp{N}_j}\right)^2= \frac{1}{M^2 \gamma^2|R|^2} \frac{1}{S_0^2}\, \sin^2 (\varphi_R+\delta_j).
\end{equation}
The sensitivity is
\begin{eqnarray}
& & \langle (\Delta \varphi_R)^2 \rangle=\sum_j \left( \frac{\partial \varphi_R}{\partial \avgp{N}_j}\right)^2 \big[ \langle N \rangle_j+\avgp{(\Delta N_T)^2}\Big] =\frac{1}{M^2 \gamma^2 |R|^2} \frac{1}{ S_0^2} \nonumber \\
& & \times \Big\{ 2S_0 \sum_j \sin^2 (\varphi_R+\delta_j) +  2|R| \gamma\,S_0 \sum_j \sin^2 (\varphi_R+\delta_j)\,\cos (\varphi_R+\delta_j)   \nonumber \\
& & + \avgp{(\Delta N_T)^2} \sum_j \sin^2(\varphi_R+\delta_j) \Big\} =\frac{1}{M^2 \gamma^2 |R|^2} \frac{1}{S_0^2} M \left\{ S_0  + \frac{\avgp{(\Delta N_T)^2}}{2} \right\}.
\end{eqnarray}
If we make the measurement $n$ times, the phase sensitivity is
\begin{equation}
\label{variance8}
\langle (\Delta \varphi_R)^2 \rangle=\frac{1}{M \gamma^2 |R|^2\, n\, S_0}\, \left[ 1+  \frac{\avgp{(\Delta N_T)^2}}{2 S_0} \right].
\end{equation}
If we make use of the visibility $V$ given by Eq. (\ref{visibility}), we can write Eq. (\ref{variance8}) as
\begin{equation}
\label{variance9}
\langle (\Delta \varphi_R)^2 \rangle=\frac{1}{M V^2 \, n\, S_0}\, \left[ 1+  \frac{\avgp{(\Delta N_T)^2}}{2 S_0} \right].
\end{equation}

\vspace{5mm}
\section{D. Characterization of the noise}
\label{App.B}

\begin{figure}[hbtp]
	\centering
\includegraphics[width=.8\linewidth]{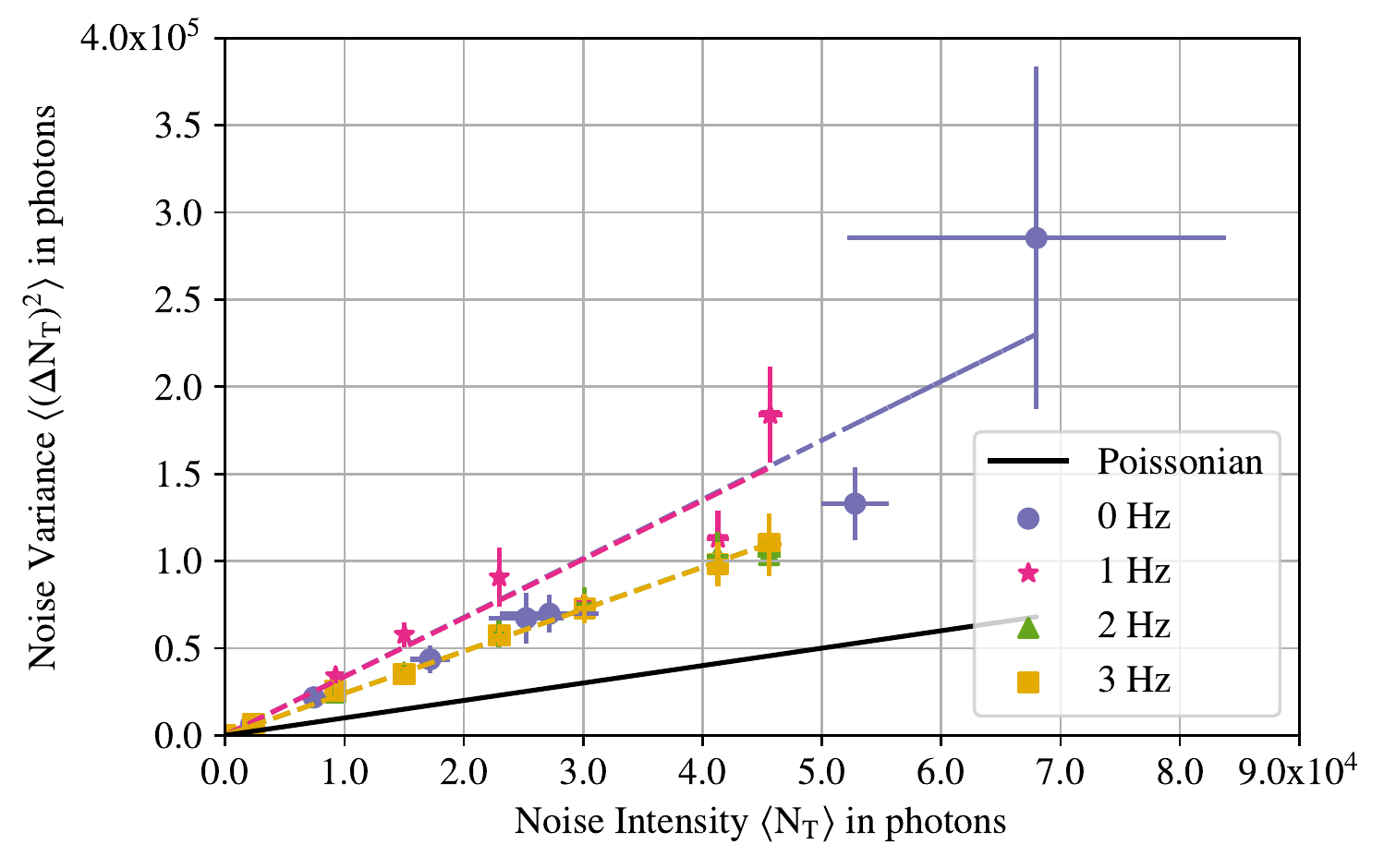}
	\caption{\textbf{Noise variance against noise intensity.} The noise properties were characterized for different configurations. For this, the mean intensity of the noise was increased while a light diffuser was rotating at different speeds. Experimental data points of the noise variances are plotted against the noise intensities. The detection window was set to be $T_D=1~\rm{s}$ and the detection area $S_D$ was $32.5 \times 32.5~\rm{\mu m}^2$. A theoretical black line represents the Poissonian case, e.g., $\avgp{(\Delta N_T)^2}=\avgp{N_T}$. In all the experiments, the noise exhibits super-Poissonian statistics.}
	\label{fig:var_vs_avg}
\end{figure}

We characterized the intensity and variance of the noise. In front of the noise source, we placed a linear polarizer to vary its intensity. We also placed a light diffuser in the path to the camera that rotates at different angular frequencies. In Fig.~\ref{fig:var_vs_avg}, we plot experimental data for the noise variances against noise intensities. The purple circle represents an angular frequency of 0 Hz. Similarly, the pink start, green triangle, and yellow square represent angular frequencies of 1, 2, and 3 Hz. A theoretical black line representing $\avgp{(\Delta N_T)^2}=\avgp{N_T}$, is also provided. All configurations of noise employed in the experiments show super-Poissonian statistics.  

\vspace{5mm}
\section{E. Signal intensity affected by noise}
\label{App.C}
\begin{figure}[hbtp]
	\centering
		 \includegraphics[width=0.8\linewidth]{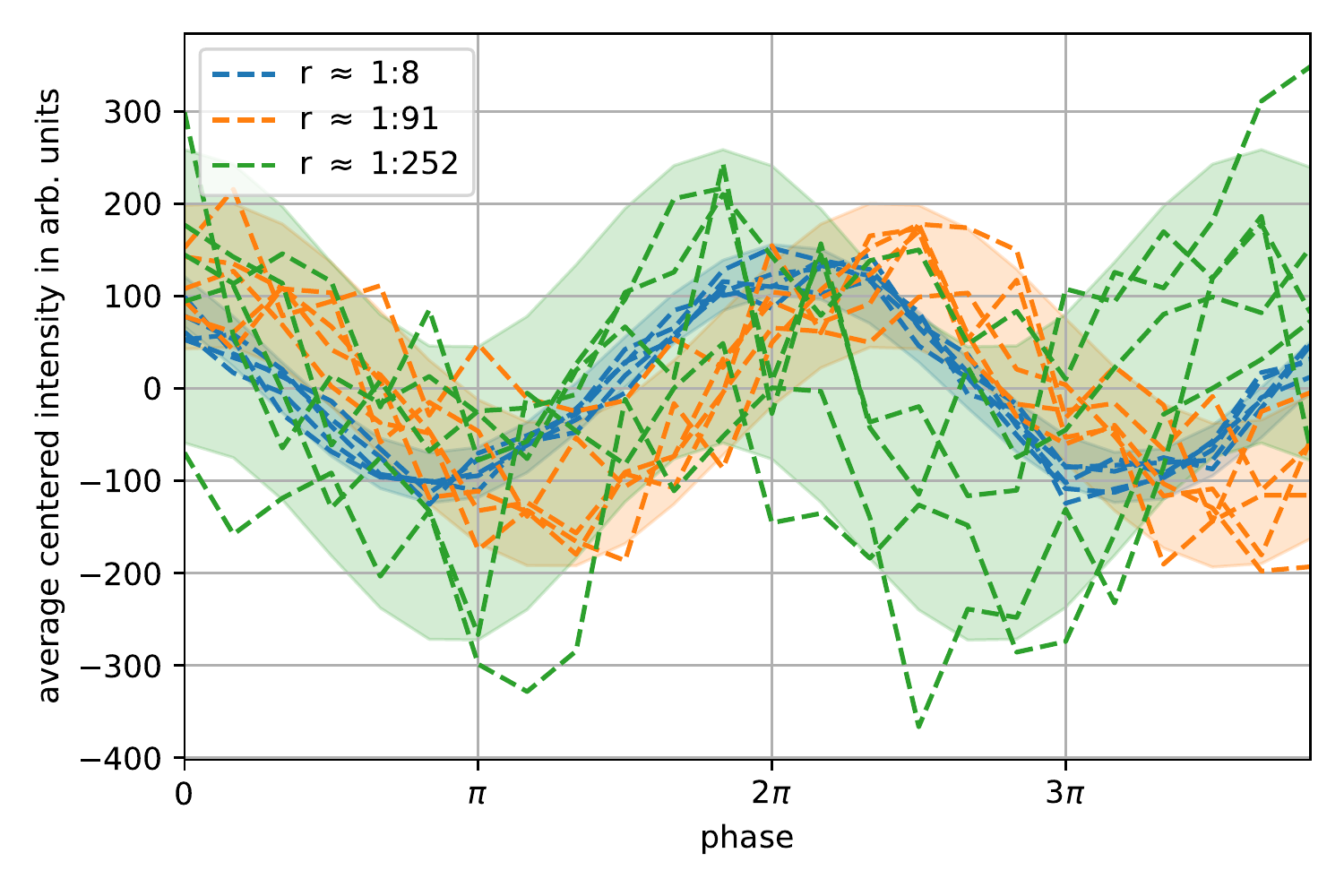}
\caption{\textbf{Signal intensity in QHUL affected by noise.} Applying a QHUL of 12 steps, we obtained the signal intensity against the phase. The experimental signal intensities are shown with dotted lines. The solid-shaded areas show the signal variances for different noise intensities. A higher noise intensity increases the signal variance. More details in the main text.}
	\label{fig:signal_superposition}
\end{figure}

In quantum holography with undetected light (QHUL), the signal intensity varies depending on the phase value. If an additional source of noise is superimposed on the camera, the signal variance increases. Figure~\ref{fig:signal_superposition} shows five QHUL measurements of 12 steps for the signal intensity collected by one pixel. We have also superimposed a noise in different ratios to the mean signal intensity. Solid-shaded areas represent the obtained signal variances. The blue area represents a ratio of $r\approx 1:8$ and resulted in a small signal variance. The orange area represents a ratio of $r\approx1:91$ and shows an increment in the signal variance with respect to the blue one. Finally, the green area represents a ratio of $r\approx1:252$, and we obtained the biggest variance for the signal photon.   

\section*{F. Statistical analysis of results presented in Fig. 5 }
\label{App.F}
In this appendix we present a further statistical analysis of the data presented in Fig.~5. These values can be found in Table~S1. For our statistical analysis we employ the so-called \textit{log-log plot} which is a technique well-known in optical coherence tomography~\cite{vinegoni2004nonlinear}. By this we obtain a linear function $Y=mX+b$, where $X=\log\{\avgp{(\Delta N_T)^2)}\}$ and $Y=\log\{\avgp{(\Delta \varphi_R)^2}\}$. $m$ is the angular coefficient that for $m=1$ represents a linear function. In Fig.~S3 are presented the fit curves, obtaining all $m$-values close to 1. This confirms that the noise and phase variances have a linear dependence.  

\begin{figure}[H]
	\centering
\includegraphics[width=.8\linewidth]{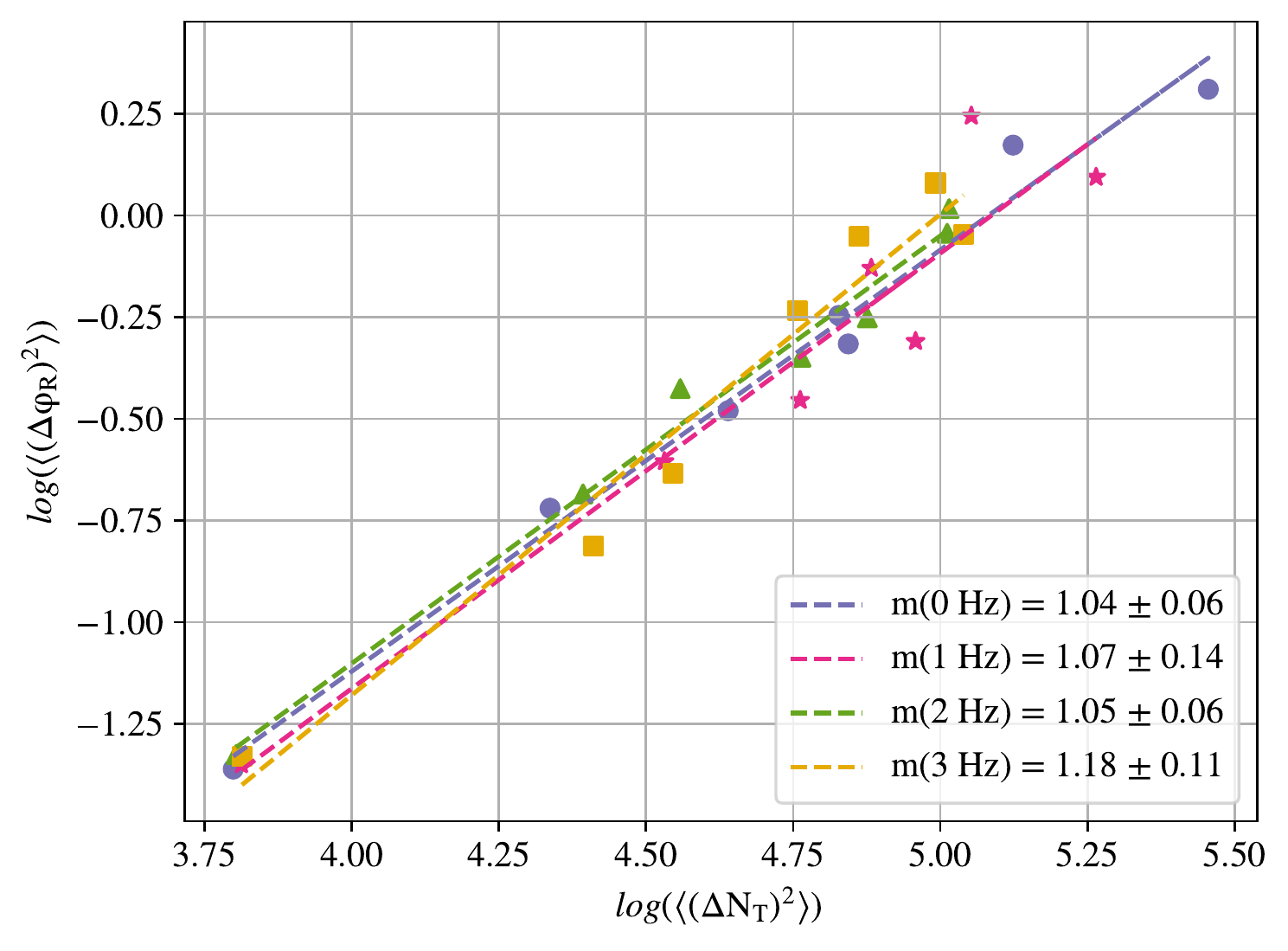}
	\caption{\textbf{Statistical analysis on the noise and phase variances}. The log-log values of the experimental data are represented by different shapes and colors for different angular speeds, see inset. A linear function $Y=mX+b$ is fit over the data points, where $m$ is the angular coefficient. By having values of $m$ near to 1 for all fit curves, we can confirm that the noise and phase variances have a linear dependence between them.}
	\label{fig:st.analysis}
\end{figure}

\begin{table}[H]
\centering
\setlength\extrarowheight{-3pt}
\begin{tabular}{||c|c|c||}
\hline
& Noise variance $\avgp{(\Delta N_T)^2}$ & Phase variance $\avgp{(\Delta \varphi_R)^2}$\\
\hline
\multirow{8}{4 em}{0 Hz} & 0 & 0.011 $\pm$ 0.005 \\
& 6292$\pm$1042 & 0.044$\pm$0.025 \\
& 21728$\pm$5609 & 0.191$\pm$0.194 \\
& 43617$\pm$7997 & 0.331$\pm$0.277 \\
& 67202$\pm$14342 & 0.568$\pm$0.390 \\
& 69767$\pm$10895 & 0.483$\pm$0.355 \\
& 132928$\pm$20766 & 1.490$\pm$0.827 \\
& 285295$\pm$ 98281 & 2.044$\pm$0.858 \\
\hline
\multirow{8}{4 em}{1 Hz} & 0 &  0.011 $\pm$ 0.005 \\
& 6498$\pm$1135 & 0.045 $\pm$0.024 \\
& 33989$\pm$4785 & 0.248$\pm$0.144 \\
& 57801$\pm$7108 & 0.351$\pm$0.297 \\
& 76327$\pm$8422 & 0.743$\pm$0.439 \\
& 90763$\pm$17137 & 0.491$\pm$0.409 \\
& 112891$\pm$16084 & 1.757$\pm$0.570 \\
& 183957$\pm$27457 & 1.243$\pm$0.733 \\
\hline
\multirow{8}{4 em}{2 Hz} & 0 & 0.011 $\pm$ 0.005 \\
& 6364$\pm$720 & 0.047$\pm$0.024 \\
& 24719$\pm$3254 & 0.207$\pm$0.186 \\
& 36157$\pm$4295 & 0.375$\pm$0.397 \\
& 57995$\pm$7749 & 0.448$\pm$0.399 \\
& 75174$\pm$9738 & 0.560$\pm$ 0.477 \\
& 102718$\pm$13822 & 0.905$\pm$0.609  \\
& 103500$\pm$ 12408 & 1.040$\pm$0.499 \\
\hline
\multirow{8}{4 em}{3 Hz} & 0 & 0.011 $\pm$ 0.005 \\
& 6511$\pm$926 & 0.047$\pm$0.023 \\
& 25715$\pm$4189 & 0.154$\pm$0.067 \\
& 35132$\pm$4263 &  0.232$\pm$0.146\\
& 57250$\pm$6479 &  0.584$\pm$0.565\\
& 72825$\pm$8410 &  0.888$\pm$0.529\\
& 98111$\pm$12569 & 1.202$\pm$0.711 \\
& 109504$\pm$18029 & 0.897$\pm$0.462 \\
\hline
    \end{tabular}
    \caption{\textbf{Detailed results presented in Fig.~5.} The noise variance errors correspond to the different values obtained at different camera pixels. The quantum holography procedure augments some errors in the phase variance because they have eluded the unwrapping script.}
    \label{tab:dist}
\end{table}